# Wide-angle and high-efficiency achromatic metasurfaces for visible light

Zi-Lan Deng,[1] Shuang Zhang,[2] and Guo Ping Wang[1,*]

[1]College of Electronic Science and Technology and Key Laboratory of Optoelectronic Devices and Systems of Ministry of Education and Guangdong Province, Shenzhen University, Shenzhen 518060, China.
[2]School of Physics & Astronomy, University of Birmingham, Birmingham B15 2TT, UK.
*Email: gpwang@szu.edu.cn

**Abstract**

Recently, an achromatic metasurface was successfully demonstrated to deflect light of multiple wavelengths in the same direction and it was further applied to the design of planar lenses without chromatic aberrations [Science, 347, 1342(2015)]. However, such metasurface can only work for normal incidence and exhibit low conversion efficiency. Here, we present an ultrawide-angle and high-efficiency metasurface without chromatic aberration for wavefront shaping in visible range. The metasurface is constructed by multiple metallic nano-groove gratings, which support enhanced diffractions for an ultrawide incident angle range from 10º to 80º due to the excitations of localized gap plasmon modes at different resonance wavelengths. Incident light at these resonance wavelengths can be efficiently diffracted into the same direction with complete suppression of the specular reflection. This approach is applied to the design of an achromatic flat lens for focusing light of different wavelengths into the same position. Our findings provide a facile way to design various achromatic flat optical elements for imaging and display applications.

Optical elements without stringent incident angle requirement are highly desired in practical imaging and display systems, such as the full-angle photographing and panoramic viewing system. There are several conventional ways for realizing wide-angle imaging systems without chromatic aberration, such as introduction of achromatic doublet or achromatic triplet, and multiple order diffractive lenses [1]. These methods, however, require cascaded multiple thick components and are therefore bulky and cumbersome.

To overcome the disadvantage of traditional optical elements, much attention has been paid to ultra-thin flat structures, known as metasurfaces, which consist of two-dimensional array of subwavelength structures [2-22] and exhibit various functionalities including wave deflection [2-7], focusing [8-10], holographic display [11-14], polarization multiplexing [15-17], and beam-shaping [18-21]. Recently, achromatization was demonstrated at multiple telecommunication wavelengths by a suitably designed metasurface composed of coupled rectangular dielectric resonators (RDRs) [23,24]. However, two important limitations have hindered the practical application of such achromatic metasurfaces [Fig. 1(a)]: 1. The achromatic behavior only exists for an ultra-narrow incident angle range (±1º) [23], that is, the metasurfaces can work well only when the incident light is from the normal direction of the metasurfaces; 2. The conversion efficiency is relatively low due to the strong specular reflection and large scattering losses [23,24], despite the use of absorption-free dielectric materials.

In this Letter, we present a wide-angle and high-efficiency achromatic metasurface [Fig. 1 (b)]. The metasurface is a combination of multiple sets of metallic nano-groove gratings, each of which has a unique period and groove height to work for a particular wavelength. It has been shown previously that the nano-groove metasurface can deflect light with very high diffraction efficiency and completely suppress the unwanted specular reflection [gray arrows in Fig. 1(b)] due to the excitation of localized surface plasmon modes [21]. As the localized plasmon modes are independent of the incident angles of the illuminating light, the enhanced diffraction efficiency exists for a very wide angular range. The operating wavelength and bandwidth of diffraction are determined by the height and width of the nano-grooves, respectively. Here a narrow-band operation (with narrow groove width) [Fig. 1(c)] is chosen for each elementary grating to avoid the crosstalk between grooves of different depths. Rigorous finite element method (FEM) results reveal that, the metasurface can efficiently diffract an incident light into the same direction for different wavelengths, even if the incident angle is varied in an ultrawide range. By modulating the elementary grating with a quadratic phase profile, one can obtain a wide-angle achromatic focusing metasurface, which can focus the incident light of different wavelengths into the same position.

For a metallic nano-groove grating at the resonance wavelength of localized plasmon mode [21], the relation between the reflection angle $\theta_r$ and incident angle $\theta_0$ is determined by,

$$\sin\theta_r = \sin\theta_0 - \lambda/p, \quad (1)$$

where, $\lambda$ and $p$ are the wavelength of incident light and period of the grating, respectively. From Eq. (1) we see that, the reflection angles are dependent on the wavelengths of light, which is the well-known chromatic dispersion existing in a common grating. If $\lambda/p$ is fixed, however, the reflection angle will be fixed for a given incident angle $\theta_0$. Therefore, in order to obtain the same diffraction angle for multiple wavelengths ($\lambda_0$, $\lambda_1$, and $\lambda_2$), one can choose multiple gratings ($g_0$, $g_1$, and $g_2$) with fixed wavelength/period ratio ($\lambda_0/p_0=\lambda_1/p_1=\lambda_2/p_2$). To ensure the enhanced diffraction in the -1st order [21], the groove heights ($h_0$, $h_1$, and $h_2$) of the multiple gratings should be suitably chosen so that the lights with wavelengths $\lambda_0$, $\lambda_1$, and $\lambda_2$ just excite the corresponding localized gap plasmon modes in the nano-grooves, respectively. By combining those elementary gratings together, we could hence expect to obtain an achromatic metasurface as schematically illustrated in Fig. 1(c).

To determine the groove height of each elementary grating, we calculated the dependence of the -1st diffraction efficiency (denoted as $R_{-1}$) of gratings upon groove height $h$ and incident wavelength $\lambda$ [Fig. 2(a)] by FEM, which is implemented by a commercial software COMSOL. The ratio between the incident wavelength and the grating period, the groove width and the incident angle are fixed in the calculations at $\lambda/p =1.1$, $w=10$nm, and $\theta_0=45°$, respectively. Silver is chosen as the grating material, with permittivity obtained by fitting the experimental data [25] to the Drude model. It is seen from Fig. 2(a) that, the spectral peak wavelength of $R_{-1}$ is linear to the groove height of the grating. When the groove height is varied from 10nm to 50nm, the wavelength of peak diffraction efficiency spans the entire visible wavelengths ranging from 400nm to 700nm. Hence we select three sets of gratings - $g_0$: $p_0$=400nm, $h_0$=15nm; $g_1$: $p_1$=500, $h_1$=29nm; and $g_2$: $p_2$=600nm, $h_2$=42nm, corresponding to wavelengths in the blue ($\lambda$=440nm), green (550nm), and red (660nm) color regions as indicated by the blue circle, green square and red triangle, respectively, in Fig. 2(a). The diffraction efficiencies of the $0^{th}$ ($R_0$) and $-1^{st}$ ($R_{-1}$) orders as well as the absorption (A) of the three gratings are shown in fig. 2(b). It is clear that the $-1^{st}$ diffraction efficiencies (solid curves) of the three gratings exhibit resonant peaks at 440nm, 550nm and 660nm, respectively. In contrast, the unwanted $0^{th}$ diffraction efficiencies (dashed curves) approach zero at the three resonant wavelengths, indicating a nearly complete suppression of the specular reflections. The absorption (represented by dot-dashed curves) also exhibits peak values at the resonant wavelengths, and the shorter the resonant wavelength, the stronger the absorption due to the stronger intrinsic absorption of metal at shorter wavelength. Nevertheless, the $-1^{st}$ diffraction efficiencies can still reach 0.78, 0.76, and 0.60 for wavelengths of 660 nm, 550 nm, and 440 nm, respectively, which are much higher than that of the recently demonstrated RDRs-based achromatic metasurface with a total theoretical transmission efficiency of 0.40 and a focusing efficiency of 0.28 [23,24].

To shed light on the origin of the enhanced diffraction efficiency of the $-1^{st}$ order, we plot the field patterns ($|H_z|^2$) at the $R_{-1}$ peak wavelengths of the three gratings in Fig. 2(c). The fields are mainly localized inside the groove regions of the gratings with a small part penetrating into the metal, which are typical field profiles of the localized gap plasmon modes. It indicates that the strong diffraction efficiencies of $R_{-1}$ are indeed caused by the excitations of the localized gap plasmon modes. Because the localized plasmon modes are independent of the incident angles of the illuminating light, the enhanced diffraction efficiencies should exist for a broad range of incident angle. Figures 2(d)-2(f) plot the $R_{-1}$ as a function of the incident angle $\theta_0$ and wavelength $\lambda$ of the illuminating light for the three gratings, respectively. The peak wavelength of $R_{-1}$ for all the three gratings remains almost constant for incident angle ranging from $10^\circ$ to $80^\circ$, which accounts for the wide-angle performance of the achromatic metasurface.

By combining three sets of the elementary gratings ($g_0$, $g_1$, $g_2$), we can construct an achromatic metasurface as shown by the bottom panel in Fig. 1(c). The resonance wavelengths of the three gratings are 440nm, 550nm, and 660nm, which correspond to blue, green and red colors, respectively. Figures 3(a)-3(c) show the FEM simulations of the combined metasurface illuminated by a Gaussian beam of blue (440nm), green (550nm), and red (660nm) color, respectively. When a Gaussian beam illuminates on the metasurface from the left side with an incident angle of $45^\circ$, the beam is reflected to the left side (negative reflection), whereas the specular reflection (to the right side) is completely suppressed. For all the three different wavelengths, the directions of the negative reflection remain the same [Figs 3(a)-3(c)]. It is because when a Gaussian with a specific wavelength illuminate the metasurface, only the gap plasmon mode in the nano-grooves that resonate at this wavelength is excited, whereas other nano-grooves do not respond to this wavelength due to large separation of the resonance frequencies and the negligible coupling between localized resonances of the nano-grooves [Fig. 2(b)]. For comparison, we also plot the case for an ordinary metallic grating illuminated by the same Gaussian beams as shown in Figs. 3(d)-3(f). The diffraction directions are apparently different for diffraction wavelengths, which is the well-known chromatic characteristic of a conventional grating.

The achromatic diffraction can exist for an ultrawide range of incident angles, as verified by the simulation for other incident angles as shown in Fig. 4. When the Gaussian beam illuminates the metasurface at $\theta_0=33°$ [Figs. 4(a)-4(c)], the beam is reflected back along its original path, because this particular incident angle satisfies the Littrow mount condition. Again, this works for all three wavelengths. The achromatic diffractions of the metasurface also exist for incident angle of 60° for the three wavelengths as shown in Figs. 4(d)-4(f). The incident beams are routed into the -1$^{st}$ diffraction order with the same diffraction angle (13.5°), while the specular reflections are completely suppressed. Therefore, the achromatic diffractions are indeed applicable for a wide incident angle range, which is highly desired for practical operations.

The wide-angle achromatic diffraction of the metasurface can be extended to the achromatic shaping of arbitrary wavefront with wide-angle and high efficiency performance. Based on the relation between the phase and grating period [21], we can modulate the periods of the three sets of elementary gratings for achieving a desired phase profile designed by holographic techniques [26,27], and then combine the modulated gratings to form an achromatic metasurface for shaping the wavefront for all three wavelengths. As a simple example, we modulate three elementary gratings with a quadratic phase profile $\phi(x) = k_i x^2 / 2f$, where $i = 0, 1, 2$ corresponds to the three different colors. Due to the different wavevector for each color, this design leads to a different gradient period profile for each of the elementary grating [21]. By carefully combining those three gradient gratings in the same way as shown in Fig. 1(c), we can obtain the metasurface as an achromatic flat lens for focusing. Figures 5(a)-5(c) show the field patterns ($|H_z|^2$) when the focusing metasurface is obliquely ($\theta_0 = 45°$) illuminated by a Gaussian beam with different wavelengths. It is observed that the reflection waves are primarily redirected into the -1st diffraction order, and are focused at approximately the same position for different colors as indicated by the dashed line for all the three wavelengths. In comparison, Figs. 5(d)-5(f) show the corresponding field patterns for a single modulated gradient grating. The incident Gaussian beam can also be focused by the single gradient grating, but the focusing point is obviously different for different wavelengths due to the intrinsic chromatic characteristic of the single grating. Therefore, the combined metasurface can indeed eliminate the chromatic aberration of a flat lens.

To demonstrate the wide-angle performance of the focusing metasurface, we also simulate the cases for other incident angles of the Gaussian beam as shown in Fig. 6. When the Gaussian beam illuminates the metasurface with incident angle 33° [Figs. 6(a)-6(c)], the beam is primarily reflected back along its original incident path, and focused in the same point (-10.7um, 16.8um) even if the color of the incident light is different. Similarly, when the incident angle is 60º [Figs. 6(d)-6(f)], majority of light can also be focused at a same point (-4.4um, 19.0um) for all the three wavelengths.

In summary, we have proposed a new type of achromatic metasurfaces with wide-angle and high conversion efficiency performances. The metasurface is constructed by integrating multiple metallic gratings with subwavelength grooves in a single surface, and the groove heights determine the wavelengths for achromatic operations. Such metasurfaces are utilized to demonstrate achromatic diffraction and focusing for visible light. The superior performance of the achromatic metasurfaces proposed here may pave the way towards practical applications in imaging and display systems.

## Acknowledgements

We thank Dr. Tao Fu for useful discussions. This work is supported by National Natural Science Foundation of China (NSFC) (Grants 11274247, 11574218, 11504243, 61501302), Natural Science Foundation of Guangdong Province, China (Grant 2015A030310400).

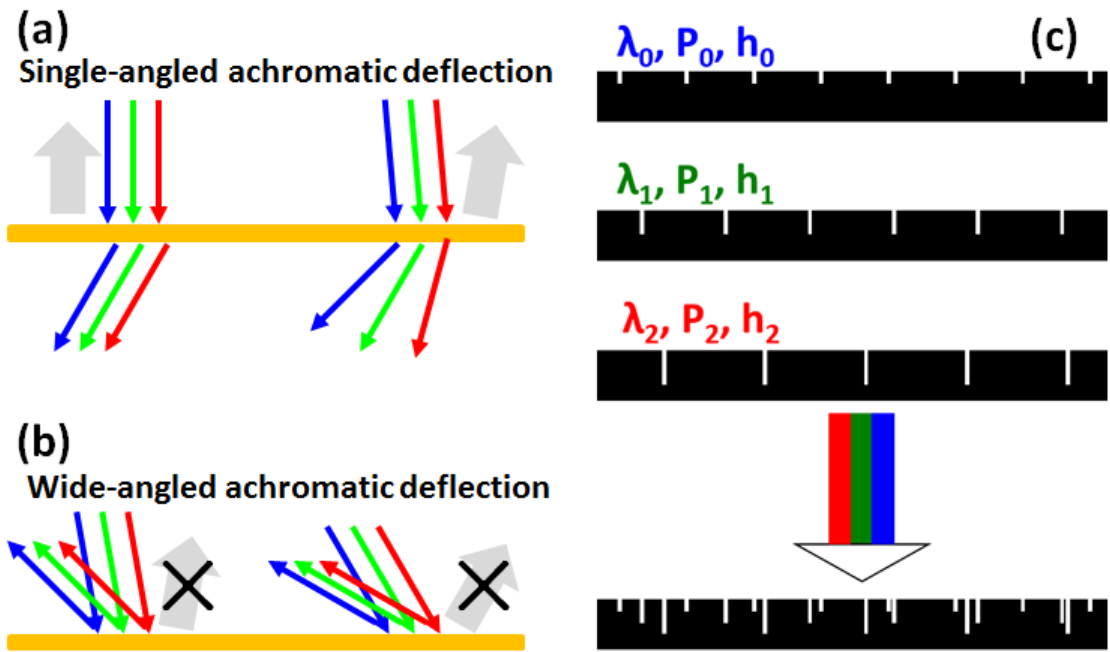

Fig. 1. (a, b) Schematic of the (a) single-angled and (b) wide-angled achromatic deflection by a metasurface (orange), respectively. (c) Schematic of formation of the ultrawide-angled achromatic metasurface by combining multiple metallic nano-groove gratings with different periods ($p_0$, $p_1$, and $p_2$) and groove heights ($h_0$, $h_1$, and $h_2$). The elementary gratings support the near-total diffraction in the -1$^{st}$ order at different wavelengths ($\lambda_0$, $\lambda_1$, and $\lambda_2$) due to the excitation of the localized gap plasmon mode in the nano-grooves with different heights ($h_0$, $h_1$, and $h_2$).

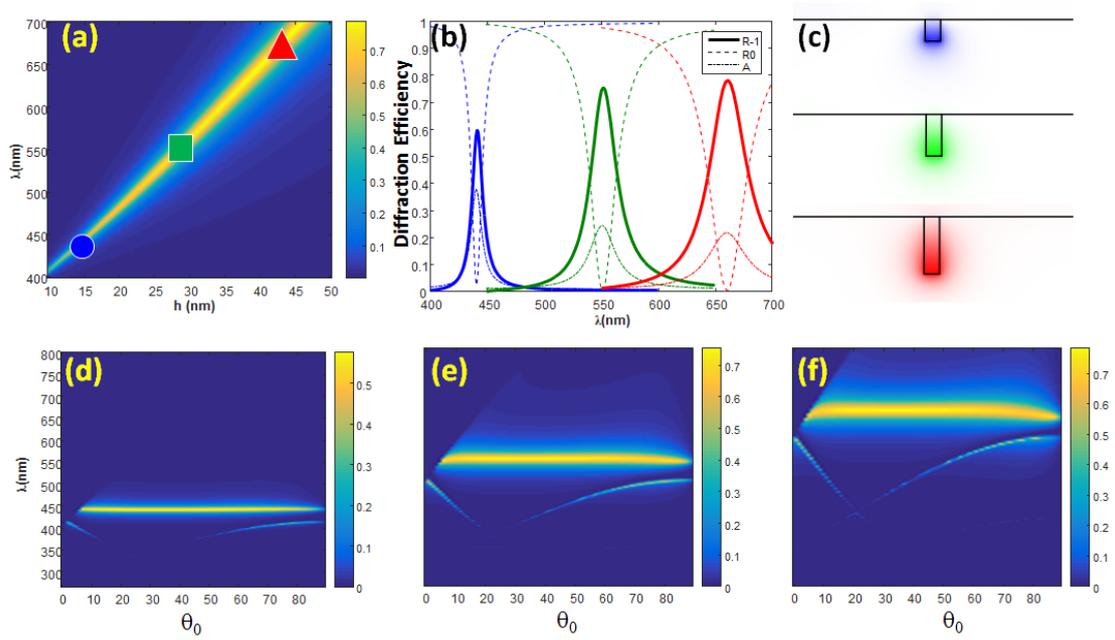

Fig. 2. (a) The -1$^{st}$ diffraction efficiency of an elementary grating for varying silt height $h$ and incident wavelength $\lambda$. The ratio of wavelength and period of the grating is fixed as $\lambda/p =1.1$. (b) The -1$^{st}$ (solid), 0$^{th}$ (dashed) diffraction efficiencies ($R_{-1}$, $R_0$) and the absorption A (dot-dashed) for three elementary gratings ($g_0$, $g_1$, and $g_2$), whose geometry parameters are indicated by blue circle, green square, and red triangle, respectively, in (a). (c) The field patterns ($|H_z|^2$) at the peak position of $R_{-1}$ of the three elementary gratings. (d-f) The -1$^{st}$ diffraction efficiency of the three elementary gratings, respectively, as a function of incident wavelength and incident angle.

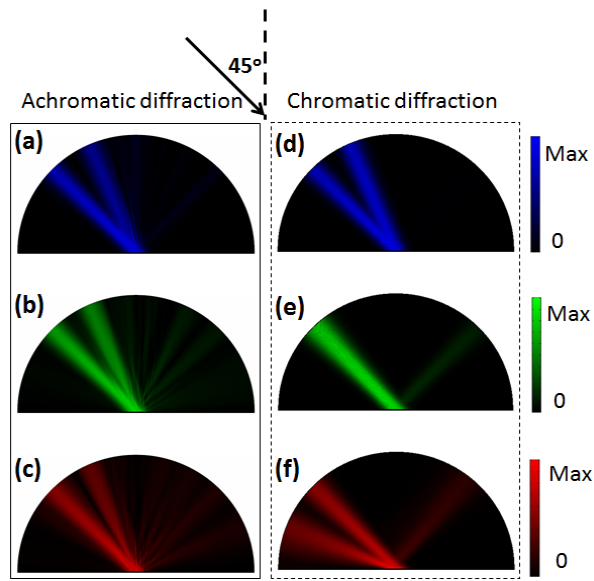

Fig. 3. (a-c) The field patterns ($|H_z|^2$) of the achromatic diffraction metasurface, which is illuminated by a Gaussian beam with (a) blue (440nm), (b) green (550nm), and (c) red (660nm) color, respectively. (d-f) The corresponding field patterns of an ordinary grating with period p=400nm for comparison. The incident angle for all cases is 45°.

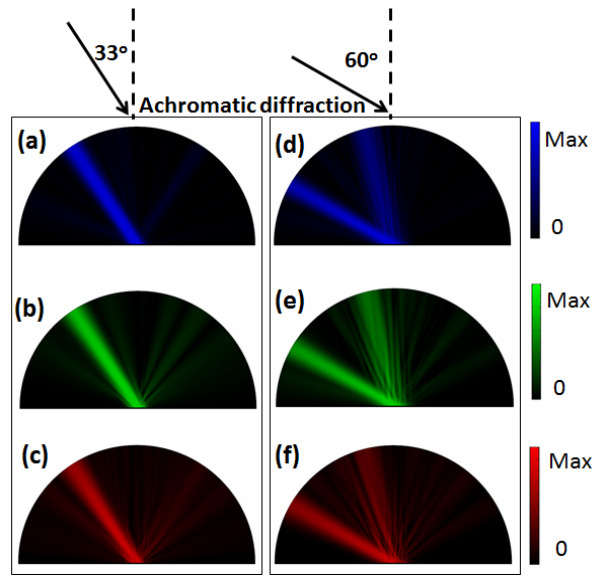

Fig. 4. The field patterns of the achromatic diffraction metasurface, which is illuminated by a Gaussian beam with (a, d) blue (440nm), (b, e) green (550nm) (c, f) and red (660nm) color, respectively, for other incident angles (a-c) 33º, (d-f) 60º, respectively.

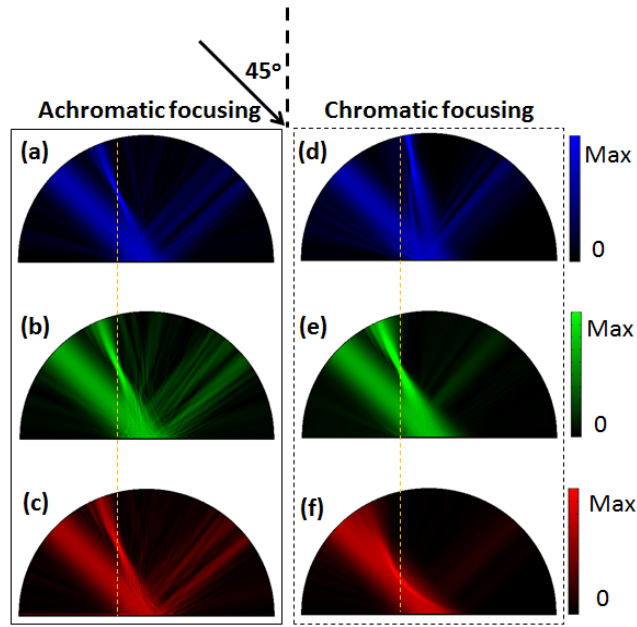

Fig. 5. (a-c) show the field patterns ($|H_z|^2$) of the achromatic focusing metasurface, which is illuminated by a Gaussian beam with (a) blue (440nm), (b) green (550nm), and (c) red (660nm) color, respectively. (d-f) show the corresponding field patterns of an ordinary gradient grating for comparison. The incident angle for all cases is $\theta_0=45°$.

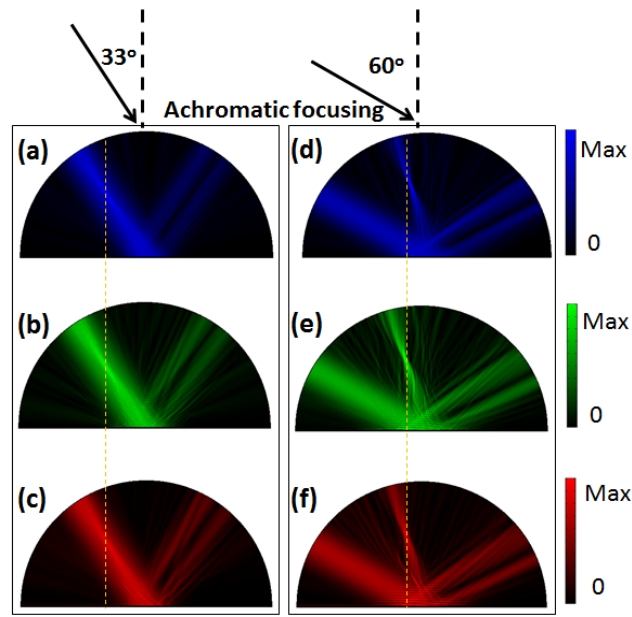

Fig. 6. The field pattern of the achromatic focusing metasurface illuminated by Gaussian beams with other incident angles (a-c) 33º, (d-f) 60º, respectively.